\documentclass[10pt,conference]{IEEEtran} 

\usepackage{tabu}                      
\usepackage{booktabs}                  
\usepackage{lipsum}                    
\usepackage{mwe}                       
\usepackage{glossaries}
\usepackage{hyperref} 

\usepackage{mathptmx}                  

\begin{document}
\title{Tables or Sankey Diagrams? Investigating User Interaction with Different Representations of Simulation Parameters}

\author{
  \IEEEauthorblockN{Choro Ulan uulu\IEEEauthorrefmark{1}\IEEEauthorrefmark{3}\thanks{Corresponding author},
  Mikhail Kulyabin\IEEEauthorrefmark{1},
  Katharina M Zeiner\IEEEauthorrefmark{1},
  Jan Joosten\IEEEauthorrefmark{1},\\
  Nuno Miguel Martins Pacheco\IEEEauthorrefmark{1},
  Filippos Petridis\IEEEauthorrefmark{1},
  Rebecca Johnson\IEEEauthorrefmark{1},
  Jan Bosch\IEEEauthorrefmark{2}\IEEEauthorrefmark{3},\\
  and Helena Holmström Olsson\IEEEauthorrefmark{4}}
  
  \IEEEauthorblockA{\IEEEauthorrefmark{1}Siemens AG, Munich, Germany\\
  Email: choro.ulan-uulu@siemens.com}
  
  \IEEEauthorblockA{\IEEEauthorrefmark{2}Department of Computer Science and Engineering,\\
  Chalmers University of Technology, Gothenburg, Sweden}
  
  \IEEEauthorblockA{\IEEEauthorrefmark{3}Department of Mathematics and Computer Science,\\
  Eindhoven University of Technology, Eindhoven, Netherlands\\
  Email: j.bosch1@tue.nl}
  
  \IEEEauthorblockA{\IEEEauthorrefmark{4}Department of Computer Science and Media Technology,\\
  Malmö University, Malmö, Sweden\\
  Email: helena.holmstrom.olsson@mau.se}
}

\maketitle

\begin{abstract}

Understanding complex parameter dependencies is critical for effective configuration and maintenance of software systems across diverse domains—from Computer-Aided Engineering (CAE) and enterprise resource planning to cloud infrastructure and database management systems. However, legacy tabular interfaces create a major bottleneck: engineers cannot easily comprehend how parameters relate across the system, leading to inefficient workflows, costly configuration errors, and reduced system trust—a fundamental program comprehension challenge in configuration-intensive software.
This research evaluates whether interactive Sankey diagrams can improve comprehension of parameter dependencies compared to traditional spreadsheet interfaces. 
We employed a heuristic evaluation using the PURE method with three expert evaluators (UX design, simulation, and software development specialists) to compare a Sankey-based prototype to traditional tabular representations for core engineering tasks.
Our key contribution demonstrates that flow-based parameter visualizations significantly reduce cognitive load (51\% lower PURE scores) and interaction complexity (56\% fewer steps) compared to traditional tables, while making parameter dependencies immediately visible rather than requiring mental reconstruction. By explicitly visualizing parameter relationships, Sankey diagrams address a core software visualization challenge: helping users comprehend complex system configurations without requiring deep tool-specific knowledge.
While demonstrated through CAE software, this research contributes to program comprehension and software visualization by showing that dependency-aware visualizations can significantly improve understanding of configuration-intensive systems. The findings have implications for any software domain where comprehending complex parameter relationships is essential for effective system use, maintenance, and evolution—from database tuning interfaces to build configuration tools and enterprise system management.

\end{abstract}


\section{Introduction}

\newacronym{kg}{KG}{Knowledge Graph}
\newacronym{llm}{LLM}{Large language Model}
\newacronym{cae}{CAE}{Computer-Aided Engineering}


Computer-aided engineering (CAE) software employs mathematical models to predict system behavior across engineering domains \cite{walker2005computer}, enabling virtual testing that reduces costs and accelerates development \cite{nvidiasimscale}, \cite{ulan2026ai}. These tools have become essential in industries from aerospace to automotive, where engineers simulate everything from aerodynamics to manufacturing quality \cite{bios12070491}, \cite{ulan2026ai}. This approach helps improve product designs and assists in resolving engineering problems. CAE encompasses the simulation, validation, and optimization of products, processes, and manufacturing tools \cite{ulan2026ai}. CAE software is heavily connected with the engineering field. It has numerous parameters users can tune to optimize their models \cite{stamatelos1999computer}. 
Modern CAE products include Siemens Simcenter suite \cite{heeds,amesim,3d,feemap}, ANSYS Mechanical \cite{ansysmec} and Altair HyperWorks \cite{altairhyper}.
They use different parameters to describe the physical processes when accurately simulating specific scenarios \cite{heeds}.

Analyzing and manipulating parameters in the engineering context is generally challenging for users that for instance are familiar with 2D sketching in CAD systems \cite{TANG2023101997}. This challenge is particularly acute for engineers who are domain experts but have limited experience with specific simulation tools - a significant user segment in industrial CAE contexts. Besides, simulation software produces large amounts of diverse data. Domain experts often struggle to explore and analyze this data because it is not integrated across various applications \cite{ziegler2020graph}. Established CAE systems typically provide comprehensive capabilities spanning multiple domains including mechanical systems, electrical systems, fluid dynamics, and thermodynamics, thereby increasing system comprehension complexity \cite{ulan2026ai}.

Conducting a high-quality experiment is a complex and challenging task \cite{POPOVICS2016257}. 
In addition to creating a precise geometric model, users must find the right parameters and accurately set the initial conditions of the experiment, a critical step in the conceptual validation of any simulation model \cite{sargent2010verification}. In some models there are many parameters, which makes it quite challenging to tune and monitor all parameters and connections between them, necessitating formal approaches like sensitivity analysis to manage this complexity \cite{Iooss2016}. 
Simulating complex models can take hours, so incorrectly set parameters from the beginning can lead to unnecessary computational and, more importantly, time costs \cite{kennedy2001bayesian}.

The simulation field is characterized by well-established ways of working, which can make introducing new methods challenging. Proficient engineers often prefer to rely on their extensive experience rather than an unknown "black box" \cite{lee2004trust}. Therefore, creating powerful interactive interfaces is crucial. Such interfaces must go beyond simply displaying data; they need to offer high degrees of interrogation, manipulation, editability, and goal expression to build trust and prove their value \cite{keim2010mastering}.
Many CAE solutions depict data in tabular representations (also spreadsheets, tables), which can be challenging to interpret because of the numerous relationships involved \cite{mirman2021graphing}.
The challenge of managing interconnected configuration parameters through isolated tables extends far beyond CAE software to enterprise systems, databases, development tools, and content creation software. This widespread reliance on tabular interfaces creates a universal problem: users must mentally reconstruct hidden dependencies between related settings, leading to errors and inefficiency \cite{LARKIN198765}. Whether configuring an ERP system, tuning database performance, or setting up engineering simulations, users face the same cognitive barriers when cause-effect relationships are obscured across multiple tables.

In our experience with simulation tools exemplifies this broader pattern—all tools use tables for parameter interaction, and users consistently struggle to understand parameter relationships. This project investigates whether alternative visualizations, specifically Sankey diagrams, could address this universal design challenge. We selected Sankey diagrams based on their existing use in engineering for flow visualization and preliminary feedback from simulation engineers who noted that parameter relationships in their mental models resembled flow patterns. Additionally, Sankey diagrams are already familiar in engineering contexts for visualizing energy and material flows, making them a natural candidate for parameter flow visualization.

To address these challenges, this research developed an interactive Sankey diagram visualization prototype and compared it to tabular representation through heuristic evaluation.
This is not to say that there are no other alternative representations but we are focusing on Sankey diagrams for this evaluation.

Based on this, the following research question is formulated:
\newline
\newline
\textit{RQ:} In what ways does visualizing engineering parameters with Sankey diagrams, as opposed to traditional tables, affect an engineer's cognitive load and in turn comprehension of parameter relationships and efficiency? 
\newline
\newline

This research directly evaluates whether Sankey diagrams can overcome the documented limitations of traditional tabular interfaces in engineering software. We conducted an expert usability review using the PURE method to compare the two representations for parameter-heavy engineering tasks, complemented by additional sessions with domain and simulation experts to validate our findings. Improving the user experience of simulation software has been shown to lead to higher satisfaction and productivity \cite{ulan2026ai}. Our results demonstrate that Sankey visualizations positively affect an engineer's ability to identify critical relationships, understand parameter impacts, and make optimization decisions more efficiently. This evaluation serves as a foundation for developing more intuitive visualization approaches for simulation tools.

The key contribution of this paper is demonstrating that flow-based Sankey visualizations significantly reduce cognitive load (51\% lower PURE scores) and interaction complexity (56\% fewer required steps) compared to traditional tabular interfaces for parameter-heavy engineering tasks. Through heuristic evaluation with domain experts, we provide empirical evidence that making parameter dependencies visually explicit—rather than requiring mental reconstruction across disconnected tables—improves program comprehension in configuration-intensive software systems. While demonstrated through CAE software, our findings have broader implications for software visualization and program comprehension: any system where users must understand complex parameter relationships (database tuning, build configuration, enterprise systems) can benefit from dependency-aware visualizations that align with users' mental models of information flow.

The remainder of the paper is structured as follows: First, the background of the challenge is described. Second, the methodology for comparing tables and Sankey diagrams is detailed. Third, the results are presented. Fourth, threads to validity are described. Fifth, the paper is summarized in the conclusion, where its future work is also stated.

\section{Background} 
\subsection{Challenges}

Research by \cite{kosmadoudi2013engineering} identifies simulation software usability challenges including high cognitive load from numerous design alternatives, user reluctance to explore efficient methods, systems that prioritize functionality over mental models, and overly complex interfaces with hundreds of menu items.

Research by \cite{lee2025new} on parametric computer-aided design tools identifies significant usability issues rooted in poor comprehension and visibility. The study found that users struggle with the high cognitive load of interpreting 3D volumes on a 2D screen. This difficulty is exacerbated by cluttered interfaces, where a high density of icons and windows makes it difficult to distinguish between functions. Furthermore, a lack of clear system feedback, such as the status of different operational tabs, adds to user confusion and inefficient workflows.

Simulation software requires users to navigate two distinct mental models: the physical system they want to analyze and the abstract computational model used for simulation. This duality is sometimes reflected in the tool architecture itself (e.g., Simcenter 3D for pre-processing and model setup versus Simcenter Amesim's distinction between physical modeling and simulation execution). While expert users seamlessly bridge these two worlds, many users struggle with this transition. Common challenges include selecting appropriate simulation parameters, defining realistic boundary conditions, interpreting results in physical context, and understanding the relationships between model assumptions and real-world behavior.

This barrier excludes the very users who could benefit most from simulation insights:
\begin{itemize}
\item Business stakeholders making strategic decisions without the technical depth to interpret simulation outputs
\item Junior engineers forced to navigate complex parameter spaces before developing the necessary intuition
\item Domain experts whose valuable knowledge is blocked by unfamiliar simulation interfaces
\item Decision-makers who must act on results they cannot independently validate.
\end{itemize}

The way simulation parameters are presented, organized, and modified fundamentally impacts tool accessibility and usability for these different user groups. 
Parameters form the primary interface for controlling CAE simulations, directly affecting accuracy, computational cost, and efficiency. As discussed in the introduction, in our experience modern models contain numerous interconnected parameters whose relationships are difficult to monitor and understand. Traditional tabular representations—still dominant in CAE tools—exacerbate this challenge by obscuring parameter relationships across multiple tables, forcing engineers to mentally reconstruct connections rather than directly perceiving them. This creates inefficient workflows and highlights the need to examine current tabular limitations and explore alternative visualization methods.

\subsection{Tabular representation}
According to \cite{furmanova2017taggle} creating complex tables is a tedious process and requires scripting skills. For instance Microsoft Excel \cite{xlsx} has rich charting operations, it provides only limited support for direct visual encoding of cells \cite{furmanova2017taggle}.
\cite{cenera2024enhancing} proposed UX guidelines to make it easier for users to create better HTML tables. Based on these guidelines they defined heuristics to determine when the usability pattern should be added to the table. This research is oriented towards improving the usability of tables that display large amounts of data. In the case of this paper, the tables are not big but are very interconnected and the information is spread across tables. 
Existing tabular representations have strength and weaknesses, but they were not made for visualizing parameter optimization. 
\cite{normalizacyjny_ergonomics_2011} has stated that the strength of spreadsheets is that they make the calculations behind a simulation transparent to the user, they allow users to participate in the full process of simulation. The disadvantage is that spreadsheets do not have visual capabilities. Researchers have tried to improve upon Excel. \cite{10.1145/237091.237097} is an interactive table viewer. It allows  the user to explore data by a combination of focus+context, a hierarchical outliner for large attribute sets and dynamic query mechanisms. But if large datasets are used with it, the data becomes unreadable. If the data is spread across different tables and is interconnected it has the same weaknesses. \cite{6875920} tried to show data spread across different tables but the user has to manually pick the blocks where the data is going to be connected.
TACO \cite{8017626} is an interactive tool that allows users to visualize the differences between multiple tables over time but it focuses on comparing versions of the same table over time. 
Therefore tabular representation should be improved to make it easier for the user to see and optimize complex engineering data relationships in ways that support decision-making.

\subsection{Sankey diagrams}
Sankey diagrams were first used to visualize the flow of energy \cite{1532152}. Sankey diagrams are flow charts, in which the width of flows is proportional to the quantity, whereas the length of a flow has no numerical meaning and can be drawn flexibly \cite{vos2022grade}. These diagrams effectively present information progressively, thereby preventing user cognitive overload from extensive parameter sets.

2023E18577 CH discloses a static, non-interactive Sankey diagram where all information is displayed simultaneously. This makes it challenging to use with a high number of parameters that are stored in simulation software. 
\cite{1532152} created a system for interactively exploring flows in Sankey diagrams. Just like 2023E18577 CH it allows users to pick different parameters to change the Sankey diagram. 

While Sankey diagrams have proven to be powerful tools for visualizing complex flow relationships, their effectiveness can be significantly enhanced through interactive implementations. Modern web-based visualization libraries, such as Plotly \cite{plotly}, provide robust capabilities for creating dynamic and interactive Sankey diagrams that allow users to explore data relationships in real-time. When combined with application frameworks like Dash—a Python-based framework designed for building analytical web applications—developers can rapidly prototype and deploy interactive visualization tools. This combination of Plotly and Dash enables the creation of responsive Sankey diagrams that support user interactions such as parameter editing, node selection, and flow highlighting, thereby addressing the limitations of static visualizations when dealing with interconnected parameter spaces commonly found in simulation software.

\section{Methodology}

The comparison between tabular representation and Sankey diagrams was performed using the PURE method (Pragmatic Usability Rating by Experts) \cite{rohrer2016practical}.

This analytical method is particularly well-suited as it allows for fast quantitative assessment of the usability of a product \cite{rohrer2016practical}, without requiring large user samples like an empirical method such as usability testing or surveys. It was therefore was selected over empirical user testing as an appropriate first-stage evaluation method in our industrial context where access to representative users is limited and costly. 
 Similarly to a heuristic evaluation, the PURE method evaluates usability by focusing on factors that are  known to affect the usability of a product. It achieves this by focusing on the in-depth cognitive load of a representative user's workflow, which is then analyzed and scored by a panel of multiple experts. 

This methodology was selected due to its capacity to minimize inter-evaluator variability affecting other analytical approaches such as cognitive walkthrough or guideline review \cite{john1997tracking}.

PURE results allow for an assessment of overall cognitive effort for the user as opposed to simply counting clicks. This is especially relevant because a single click does not reflect the cognitive load involved in performing that click. The method's validity and utility have been demonstrated in both academic and industrial contexts. For instance, in its foundational case study, \cite{rohrer2016practical} successfully used PURE to generate task usability ratings for three software products, demonstrating that the method's results correlated with metrics from traditional, large-scale usability tests. Other studies that have used the PURE method are e.g. \cite{parameswari2024task} to assess the usability of voice user interface to command exoskeleton domains and  \cite{khanvilkar2022digidrive} to assess the usability of a mobile application of a driving school. 

PURE is an analytical expert evaluation method where trained evaluators systematically assess cognitive load across five dimensions. A panel of experts independently analyzed both interfaces, then collaboratively assigned PURE scores through structured discussion. To ensure assessments reflected realistic usage patterns, experts referenced video recordings of representative task completion workflows during their analysis.

The PURE analysis was conducted through expert cognitive walkthroughs of representative engineering tasks. A panel of three expert  evaluators independently assessed both interfaces by systematically working  through each task while rating the cognitive load required at each step. The evaluators leverage their combined expertise in UX design, engineering domains, and software implementation to predict the cognitive demands a typical engineer would experience. This multi-evaluator approach reduces individual bias and increases reliability compared to single-evaluator heuristic methods \cite{10.1145/142750.142834}.

\subsection{Task description.}
For a PURE analysis the overall workflow is split into tasks (e.g. add parameter) and task steps (i.e. all the steps needed to complete a given task - e.g. enter information for parameter x and click save).
The tasks underlying this analysis were to create and edit a model with both global and local parameters, which are described in the Appendix \ref{append}. Global parameters can be referenced in a local parameter. Users were asked to (i) add new global parameters, (ii) change global parameters, (iii) change how they were used in a local parameter. Users were supplied with multiple tables. (i) tables with global parameters, (ii) tables with local parameters. Global parameters were often referenced in a local parameter. For the PURE evaluation, each of these tasks was broken down into its constituent steps. 

To illustrate this process, consider Task 3: Change the formula for a local parameter that uses a global parameter.

In the tabular interface, completing this task required the following discrete steps, each of which was individually rated for cognitive load:
\begin{enumerate}
    \item Navigate to "Car 1" spreadsheet 
    \item Check calculation for "Spring deflection"
    \item Navigate to "Car 2"
\item Check calculation for "Spring deflection"
\item Navigate to "Car 3"
\item Check calculation for "Spring deflection"
\item Change the calculation for "spring deflection"
\item Navigate to "Car 4"
\item Check calculation for "Spring deflection"
\end{enumerate}
With this detailed breakdown the evaluator can pinpoint where exactly user effort is expected.
A similar step-by-step breakdown was created for the Sankey diagram interface to ensure a direct and fair comparison.

\subsection{Target User Profile }

For this heuristic evaluation the target audience was engineers with subject matter expertise but limited expertise in a given modeling tool. Subject matter expertise refers to a comprehensive understanding of the engineering domain being simulated. In the context of this evaluation's tasks (see Appendix \ref{append}), this means the participant understands the physical principles at play—such as the relationships between mass, gravity, force, and spring stiffness—independent of any software interface. They know what they need to model. Limited tool expertise means the participant is not a "power user" of the specific software being evaluated. While they may have general computer literacy, they are not expected to have memorized the specific workflows, menu locations, or interaction patterns for manipulating parameters within the tested interfaces (neither the spreadsheet nor the Sankey prototype). They do not yet have an established routine for how to perform the tasks in this particular tool. This group was chosen because it allows us to compare the performance of tables and Sankey diagrams without experience with either adding a confound and in our experience represents a significant user segment in industrial CAE contexts.

To validate the task design and ensure both interfaces could support the required workflows, we conducted a pilot session with an engineer who matched the target user profile—familiar with the subject matter but not experienced with either Excel-based parameter management or Sankey diagrams. This pilot session was video recorded and served as the empirical reference material for the subsequent expert evaluation.

\subsection{Stimuli}

\subsubsection{Spreadsheet}
\begin{figure}
    \centering
    \includegraphics[width=1\linewidth]{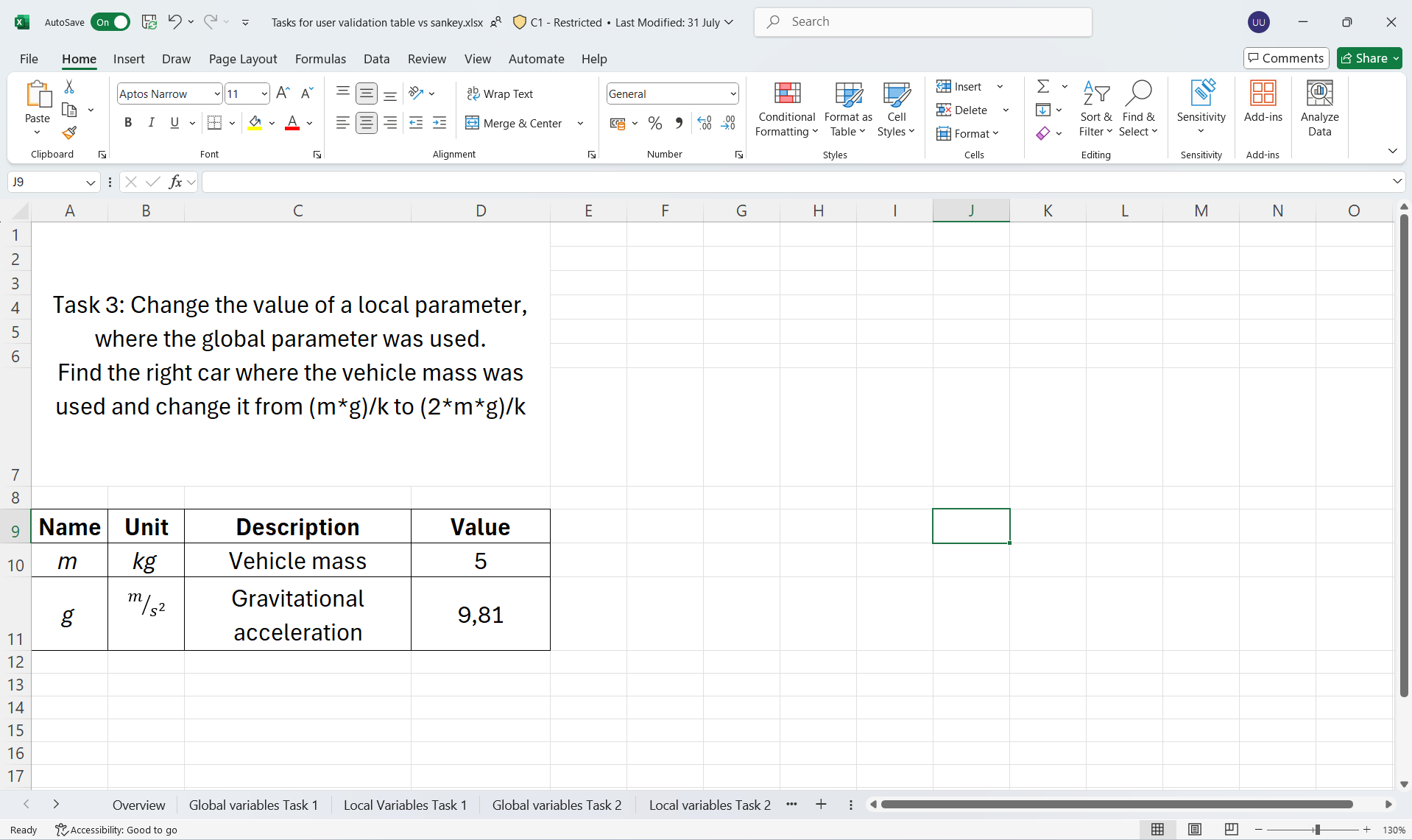}
    \caption{Microsoft Excel UI with task 3: Change the value of a local parameter, where the global parameter was used}
    \label{fig:exel}
\end{figure}

As seen in Fig. \ref{fig:exel} the tasks are spread across spreadsheets. The user has to navigate through them to see how global parameters are defined and how they were used in local parameters. 

\subsubsection{Sankey diagram}
  
Sankey diagrams show how tables are interconnected. The implementation provides the following capabilities:
\begin{enumerate}
    \item \textbf{Parameter manipulation}. Users can add and change parameters directly within the visualization
    \item \textbf{Visual relationship mapping}. When tables are connected, changing one parameter updates related values, with connections shown through links
    \item \textbf{Dynamic flow width}. Link widths change proportionally with parameter values
    \item \textbf{Guided interaction}. The system shows information step-by-step through interactive pathways
\end{enumerate}

\begin{figure}
    \centering
    \includegraphics[width=1\linewidth]{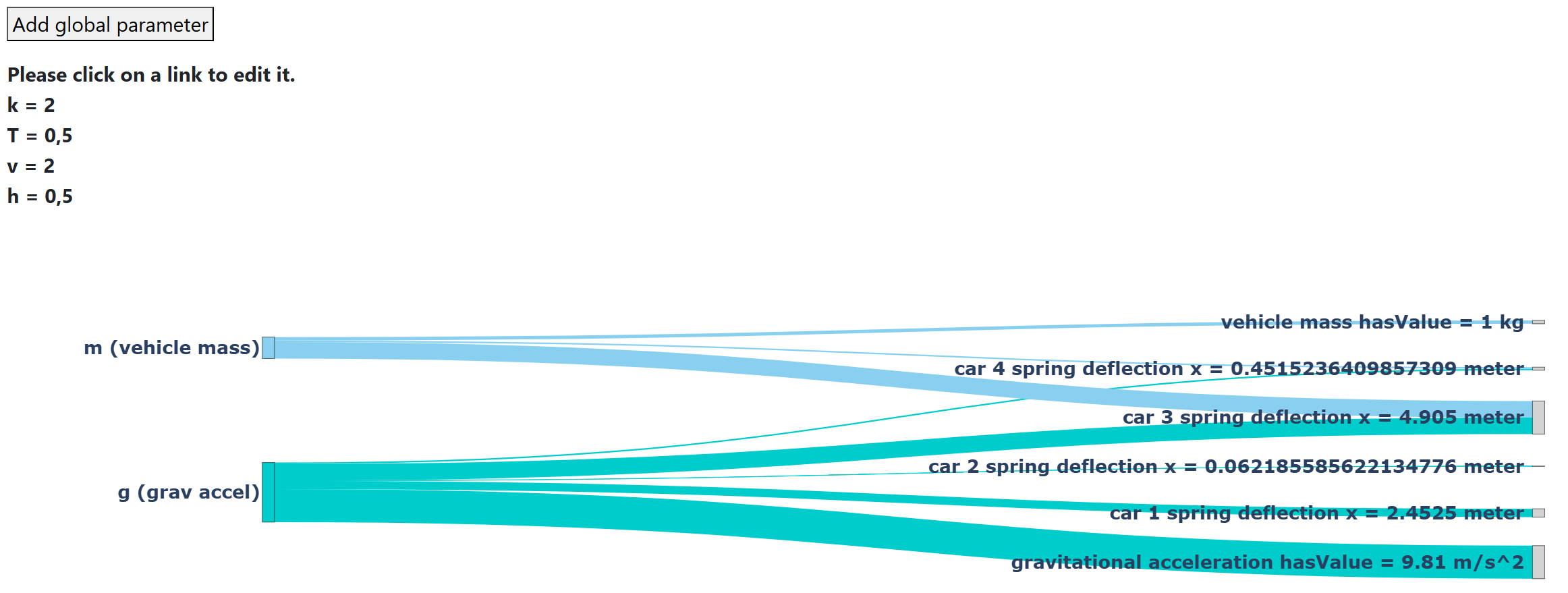}
    \caption{Sankey Diagram. The links between nodes reflect how global parameters influence local parameters. The width of links changes depending on the values}
    \label{fig:sankey}
\end{figure}

Figure \ref{fig:sankey} illustrates how this approach differs from traditional spreadsheets. Where Excel requires users to mentally trace relationships across cells and sheets, the Sankey diagram explicitly shows how global parameters flow into and influence local parameters.

\subsection{Evaluation}

The PURE evaluation proceeded in two phases:

\textbf{Phase 1: Individual Assessment.} Each of the three expert evaluators independently reviewed the video recording of the pilot session. While watching, they rated each task step from the perspective of the target user profile, assigning scores based on predicted cognitive load:

\begin{itemize}
    
\item One: the step can be accomplished easily by the target user, due to low cognitive load or because it’s a known pattern, such as the acceptance of a terms-of-service agreement. 
\item Two: the step requires a notable degree of cognitive load (or physical effort) by the target user, but can generally be accomplished with some effort.
\item  Three: the step is difficult for the target user, due to significant cognitive load or confusion. Some target users would likely fail or abandon the task at this point.
Following the ratings the evaluators discussed their individual ratings to arrive at a consensus on which rating given to each task.
\end{itemize}

\textbf{Phase 2: Consensus Discussion.} Following individual ratings, the evaluators convened for a collaborative session to discuss their assessments and arrive at consensus ratings for each task step.

The three expert evaluators brought complementary perspectives:
\begin{itemize}
    \item UX expert. Has over 10 years of experience in working in the field of user experience. Helped set up the heuristic evaluation according to PURE. 
    \item Engineering expert. Has worked in the engineering field for many years. Has experience with tabular representation in the engineering context.
    \item The developer of the Sankey diagram. Software developer that has implemented the Sankey representation and is leading the project.
\end{itemize}

To ensure consistency among the three expert evaluators, inter-rater reliability of the initial ratings was assessed using Cohen's kappa. The evaluation yielded kappa = 0.57.
While this level of agreement is moderate, it is not a problem for the overall PURE results, because the second part of the rating session arrives at a consensus rating, meaning the IRR of the final scores is 1. The initial IRR, however, suggests that before the final rating session the different evaluators were using slightly different rating schemes.

\section{Results and Discussion}

\subsection{Quantitative Performance Analysis}

The PURE evaluation revealed differences between tabular and Sankey diagram representations across all three tasks. Figure \ref{fig:pureres} summarizes the key findings.
The two approaches were compared descriptively.

The PURE evaluation revealed an important insight: step count alone does 
not determine interface complexity. 

\textbf{Task 1: Adding Global Parameters.}
Despite requiring more task steps (6 vs. 4), the Sankey diagram achieved a lower overall PURE score.  This apparent contradiction—fewer steps yielding lower scores—merits explanation. The tabular interface received predominantly "yellow" ratings (score 2) for all steps except the first, resulting in a total score of 7 (Fig. \ref{fig:pureres}). In contrast, the Sankey interface received all "green" ratings (score 1), totaling 6 points (Fig. \ref{fig:pureres}). This indicates that although Sankey approaches require additional steps, each step demands reduced cognitive effort (all rated 'green') compared to tables (predominantly 'yellow'). A simple click-counting heuristic would suggest the table is superior. This counterintuitive result demonstrates that step count or click-counting alone do not determine interface complexity because the cognitive load per step is equally important.

For our target users — engineers with domain knowledge but limited tool experience — this finding is particularly significant. While experienced users may have memorized the spreadsheet locations and navigation patterns, less-experienced users must actively search and verify each step, making the cognitive load of each "yellow-rated" table interaction substantially higher. The Sankey diagram's visual guidance reduces this burden by making the workflow self-evident, requiring less tool-specific knowledge to complete the task successfully.

This demonstrates PURE method value in capturing both interaction quantity and individual interaction mental load. The Sankey interface guides users through more steps, but each step is intuitive and 
requires minimal mental effort.

\textbf{Task 2: Editing Global Parameters.}
Both interfaces presented challenges in this task, as evidenced by yellow ratings. However, the Sankey diagram outperformed tables with a 55\% lower PURE score (10 vs. 22). The initial difficulty with the Sankey interface stemmed from the interactive link elements used to interact with the parameters. 

\textbf{Task 3: Editing Local Parameters.}
The Sankey diagram continued to demonstrate superior performance, requiring fewer cognitive resources to complete the task. The visual representation of parameter relationships enabled more efficient navigation and modification compared to searching through multiple spreadsheets and cells.

\subsection{Qualitative Observations}

The qualitative analysis revealed why the Sankey approach addresses fundamental challenges identified in prior CAE usability research and outperformed traditional tables from a cognitive alignment perspective. Expert feedback highlighted that the visualization structure mirrors engineers' mental models of parameter relationships: \textit{"The Sankey diagram reflects the structure you have in your head, but visualized."} This directly addresses \cite{kosmadoudi2013engineering}'s observation that existing systems fail to reflect users' mental states during model development, prioritizing functionality over cognitive alignment.

The consolidated Sankey view directly tackles the "overly complex interfaces with hundreds of menu items" problem documented by \cite{kosmadoudi2013engineering}. Where traditional tabular approaches force users to navigate multiple spreadsheets—exacerbating the cognitive load issues identified by \cite{lee2025new} regarding cluttered interfaces and poor visibility—the Sankey diagram presents all parameter relationships in a single, coherent view. Parameter dependencies that remain hidden across multiple tables become immediately visible through explicit flow connections, eliminating the need to mentally reconstruct relationships that \cite{furmanova2017taggle} noted requires scripting skills to even create in traditional tools.

The top-down flow structure naturally represents how global parameters influence local ones, directly addressing the dual mental model challenge we identified: bridging the physical system users want to analyze and the abstract computational model. This visual explicitness reduces the barrier for the excluded user groups we highlighted—business stakeholders, junior engineers, and domain experts—by making parameter propagation immediately observable rather than requiring deep technical knowledge to trace through spreadsheet formulas.

The Sankey approach counters \cite{kosmadoudi2013engineering}'s finding that "users are often not motivated to explore and find more efficient methods" by making parameter relationships self-evident. Unlike the static Sankey implementations in prior work (2023E18577 CH) or comparison-focused tools like TACO \cite{8017626}, our interactive implementation allows real-time parameter editing with immediate visual feedback through varying flow widths. This addresses \cite{lee2025new}'s identified need for clear system feedback and reduces the "lack of visual capabilities" limitation that \cite{normalizacyjny_ergonomics_2011} noted in traditional spreadsheets.

The measured 51\% reduction in PURE scores (indicating lower cognitive load) and 56\% fewer required interactions provide empirical evidence that Sankey diagrams overcome the usability barriers documented in prior research. Where \cite{10.1145/237091.237097}'s interactive table viewer became unreadable with large datasets and \cite{6875920}'s approach required manual connection of data blocks, our Sankey implementation automatically visualizes interconnected parameters across what would traditionally be multiple tables. This substantially improves accessibility for our target demographic—engineers with strong domain expertise but limited tool-specific familiarity—by removing the cognitive barriers that \cite{kosmadoudi2013engineering} identified as preventing users from learning alternative design methods.

The convergence of qualitative feedback with quantitative metrics suggests that Sankey diagrams address multiple documented CAE usability challenges simultaneously: they reduce interface complexity, align with mental models, improve parameter relationship visibility, and lower barriers to effective system utilization. This positions flow-based visualization as a fundamental improvement over traditional tabular approaches for parameter configuration in simulation software.

\subsection{Efficiency Metrics}
\begin{figure}
    \centering
    \includegraphics[width=1\linewidth]{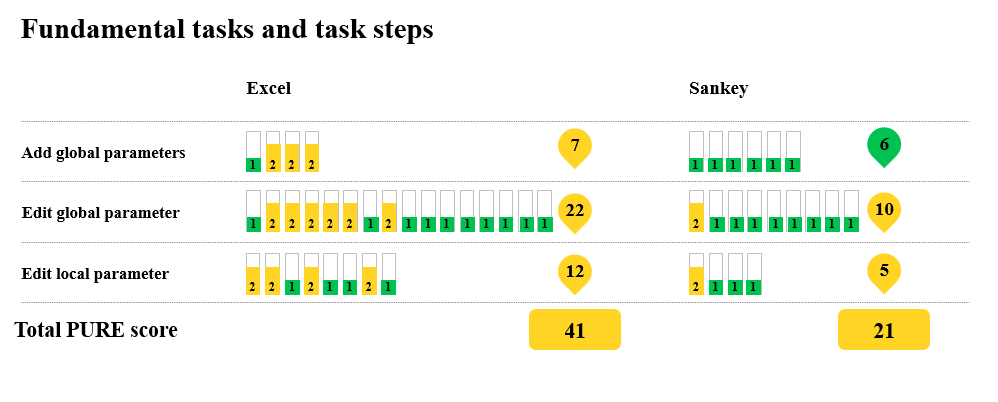}
    \caption{PURE scores. Sankey diagrams have a lower PURE score in comparison to Excel spreadsheets}
    \label{fig:pureres}
\end{figure}

\begin{itemize}
    \item Sankey-based interfaces require fewer steps to complete equivalent tasks (Fig. \ref{fig:pureres}), which may reduce likelihood of task abandonment;
    \item Sankey diagrams have a lower PURE score (Fig. \ref{fig:pureres});
\end{itemize}

This reduction in mechanical interaction, combined with the 51\% improvement in PURE scores, demonstrates that Sankey diagrams offer both cognitive and mechanical efficiency advantages.

\subsection{Use Case Scenarios}
\subsubsection{Showing and changing the global and local parameters.}
The Sankey representation facilitates how users add, change, and remove global parameters or modify the values of local parameters. In the traditional interface of the simulation software Simcenter Amesim which uses spreadsheets for parameter interactions, these operations require 9 clicks to complete, but when performed through a Sankey-based interface, the number of clicks reduces to 4. This represents a 56\% reduction in interaction steps. Based on PURE as shown the scores have been lowered by 51\% in Sankey diagrams in comparison to spreadsheets.

\subsubsection{Parameter exploration and comprehension.}
Based on the results users are able to explore and understand simulation parameter relationships more intuitively using Sankey-based visualization compared to traditional tabular representations. The flow approach allows users to focus on relevant parameters of components without becoming overwhelmed by the entire tabular representation.

\section{Threats to Validity}

\subsection{Internal Validity}
\textbf{Evaluator Bias.} The three evaluators included the developer of the Sankey prototype, which introduces potential bias toward favorable ratings of the Sankey interface. We mitigated this through: (1) using the structured PURE framework with explicit rating criteria and (2) requiring consensus among all three evaluators including independent UX and engineering experts.

The moderate inter-rater agreement (kappa = 0.57) before consensus discussion suggests evaluators were not uniformly biased, as systematic bias would likely produce higher initial agreement. Nevertheless, future evaluations should include fully independent evaluators unfamiliar with either interface design.

\textbf{Analytical Evaluation Approach.} This study employed an analytical evaluation method rather than an empirical user study. We conducted a heuristic analysis after observing one representative participant matching our target user profile, which is appropriate for this type of expert-based evaluation where the focus is on identifying usability issues through systematic inspection rather than statistical generalization. The PURE method's validity for such analytical approaches has been demonstrated with small samples \cite{rohrer2016practical}. Each task was performed once per interface, capturing the realistic initial user experience that our target users—engineers with domain expertise but limited tool experience—would encounter. This analytical approach follows established practice for early-stage interface evaluation \cite{10.1145/97243.97281}, where expert analysis of representative user behavior identifies fundamental usability issues before resource-intensive empirical studies with larger participant samples.

\subsection{External Validity}
\textbf{Task Representativeness.} The three tasks were designed to represent common parameter management operations in CAE software. While they cover core workflows (adding, editing global and local parameters), they may not capture all possible parameter manipulation scenarios. The tasks were validated through pilot sessions with domain experts to ensure relevance to real engineering work.

\textbf{Domain Specificity.} Although we position findings within broader configuration management contexts, empirical validation occurred only within the CAE engineering domain. Claims about applicability to ERP systems, database configuration, or other domains remain theoretical and require future validation.

\textbf{Scalability.} The evaluation used a moderate number of parameters. Performance with significantly larger parameter spaces (hundreds or thousands of parameters) remains unknown and represents an important area for future investigation.

\textbf{Alternative Visualization Approaches.} This study compared Sankey diagrams against current tabular practice in CAE software. While other visualization approaches for parameter dependencies could exist (e.g., dependency graphs, network diagrams), our evaluation establishes that flow-based visualization successfully addresses the cognitive load challenges of current tabular interfaces. Future work should compare different dependency visualization approaches to identify optimal solutions for specific contexts.

\subsection{Construct Validity}

\textbf{PURE Score Interpretation.} While PURE scores provide quantitative comparison, they represent expert predictions of cognitive load rather than direct measurement. The quantitative scores were triangulated with qualitative observations from verbal protocols and behavioral patterns captured in video recordings to strengthen the validity of findings.

\textbf{Target User Group Selection.} This evaluation deliberately focused on engineers with domain expertise but limited tool-specific experience. This user segment represents a critical validity consideration: if Sankey-based visualizations demonstrate cognitive load reduction for novice tool users, the benefits likely extend to expert users as well, since interfaces that simplify parameter relationship comprehension for less-experienced engineers typically maintain or improve usability for experts. Conversely, evaluating only expert users might mask usability barriers that prevent broader adoption.

Focusing on this group strengthens the external validity of findings by addressing the user segment most likely to encounter cognitive barriers with traditional tabular interfaces. Domain experts with limited tool experience represent both a substantial portion of potential CAE software users and the population most affected by interface complexity. Demonstrating effectiveness for this group provides evidence that Sankey visualizations address genuine usability challenges rather than merely offering aesthetic alternatives for already-proficient users.

\subsection{Reliability}
The moderate initial inter-rater agreement (kappa = 0.57) before consensus indicates some inconsistency in how evaluators initially applied the PURE rating scale. The structured consensus discussion resolved these differences, but highlights the importance of clear rating criteria and calibration among evaluators in future applications of this method.

\section{Conclusion}

This paper investigated how visualizing engineering parameters with Sankey diagrams, as opposed to traditional tables, affects an engineer's efficiency, cognitive load, and comprehension. The results demonstrate that Sankey diagrams improved user experience by directly addressing the core weaknesses of tabular formats. The findings from our PURE analysis provide answers to the research question.

Key findings reveal that Sankey diagrams significantly reduce cognitive load by making parameter dependencies visually explicit, enhance comprehension through immediate visual feedback, and increase task efficiency with 51\% lower PURE scores and 56\% fewer required interactions. The visual clarity particularly benefits non-experts by aligning with their conceptual understanding, though experienced users who have internalized table-based workflows showed mixed responses.

These findings extend beyond CAE software. The core insight—that explicitly visualizing parameter relationships improves usability—applies to any domain managing interconnected configurations: ERP systems could reduce misconfiguration errors, database tuning interfaces could clarify performance parameter interactions, and development tools could better communicate build configuration dependencies. As software complexity increases across all domains, the need for alternatives to isolated tabular interfaces becomes critical.

In conclusion, although demonstrated through CAE case evaluation, this research addresses fundamental configuration interface design limitations across software domains with particular benefits for users possessing domain expertise but limited tool-specific experience. By making relationships explicit and providing immediate visual feedback, Sankey diagrams offer a promising alternative to lower the barrier to entry wherever parameter interdependencies create complexity.

\subsection{Future Work}

This evaluation opens several research directions for advancing flow-based parameter visualization in software systems.

While our findings suggest broad applicability to configuration-intensive systems, empirical validation across other domains—such as database configuration, ERP systems, and build tools—would establish whether the observed cognitive load reductions generalize beyond engineering contexts. Controlled studies comparing Sankey visualizations to traditional interfaces in these domains represent a critical next step for confirming the universality of our findings.

Our evaluation employed tasks with moderate parameter quantities. Future research should systematically investigate how Sankey visualizations perform with significantly larger parameter spaces containing hundreds or thousands of interconnected parameters. This includes exploring hybrid approaches that combine tabular and flow-based representations, potentially leveraging hierarchical visualization techniques or adaptive interfaces that adjust based on parameter complexity.

This heuristic evaluation captured expert first impressions, but longitudinal studies tracking engineers over extended periods would reveal whether initial cognitive load reductions translate into sustained productivity gains in real-world workflows and how learning curves differ between traditional and flow-based interfaces.

\section{Acknowledgments}
The authors gratefully acknowledge the support provided by the Software Center. The authors also thank their colleagues and supervisors within Siemens for valuable discussions and feedback during this research. Special thanks to Kai Liu from the product software team for his contributions and support.

\bibliographystyle{IEEEtran}

\bibliography{template}

\section*{Appendix} 
\label{append}

Because in this paper the tabular representation and Sankey diagrams are compared, and because the relationships between global and local parameters are important in the engineering context, the system description and tasks are as follows: 

\subsection*{System Description:}
The model represents a simple stationary vehicle on an ideal, vertical spring. There is no movement and no damping – it is a static state. The vehicle's weight pushes the spring downward. The spring has a locally defined stiffness (local parameter). The goal is to investigate the influence of the vehicle mass (global) on the spring deflection (locally calculated).
\subsection*{Tasks}
\begin{enumerate}
    \item \textbf{Add global parameters.} Add the global variables mass (m) and gravitational acceleration (g) to the simulation model and calculate the gravitational force (Fg). Use g=9.81 $m/s^2$ as a global constant if needed
    \item \textbf{Change Global Value.} Change the value of the global variable Vehicle Mass (m) in multiple steps (e.g., 5kg, 10kg, 15kg, etc.) and simulate the spring's behavior regarding spring deflection (x)
    \item \textbf{Change the value of a local parameter, where the global parameter was used.}
Find the right car where the vehicle mass was used and change it from 
\begin{equation}
    \frac{m\cdot g}{k} 
\end{equation}
to
\begin{equation}
    \frac{2\cdot m\cdot g}{k}
\end{equation}
\end{enumerate}

\end{document}